\documentclass[preprint,showpacs,superscriptaddress,nofootinbib]{revtex4}
\usepackage[dvips]{graphicx}
\usepackage{amsmath,amssymb}

\newcommand{\lromn}[1]{\uppercase\expandafter{\romannumeral#1}}

\newcommand{\av}[1]{\langle #1 \rangle}

\newcommand{\MNS}{{\text{MNS}}}

\newcommand{\eV}{{\text{eV}}}

\newcommand{\GeV}{{\text{GeV}}}
\newcommand{\TeV}{{\text{TeV}}}
\newcommand{\BR}{\text{BR}}

\newcommand{\meg}{\mu \to e \gamma}

\newcommand{\BL}{{\text{B$-$L}}}
\newcommand{\DM}{{\text{DM}}}
\newcommand{\EM}{{\text{EM}}}

\newcommand{\pell}{{\ell^\prime}}

\begin{document}

\preprint{UT-HET 060}

\title{
 TeV-Scale Seesaw with Loop-Induced Dirac Mass Term
and Dark Matter from $U(1)_\BL$ Gauge Symmetry Breaking
}

\author{Shinya Kanemura}
\email{kanemu@sci.u-toyama.ac.jp}
\affiliation{
Department of Physics,
University of Toyama, Toyama 930-8555, Japan
}
\author{Takehiro Nabeshima}
\email{nabe@jodo.sci.u-toyama.ac.jp}
\affiliation{
Department of Physics,
University of Toyama, Toyama 930-8555, Japan
}
\author{Hiroaki Sugiyama}
\email{hiroaki@fc.ritsumei.ac.jp}
\affiliation{
Department of Physics,
Ritsumeikan University, Kusatsu, Shiga 525-8577, Japan
}


\begin{abstract}
 We show a TeV-scale seesaw model
where Majorana neutrino masses,
the dark matter mass, and stability of the dark matter
can be all originated from the $U(1)_\BL$ gauge symmetry.
 Dirac mass terms for neutrinos are forbidden
at the tree level by $U(1)_\BL$,
and they are induced at the one-loop level
by spontaneous $U(1)_\BL$ breaking.
 The right-handed neutrinos can be naturally at the TeV-scale or below
because of the induced Dirac mass terms with loop suppression.
 Such right-handed neutrinos
would be discovered at the CERN Large Hadron Collider~(LHC)\@.
 On the other hand,
stability of the dark matter is guaranteed
without introducing an additional $Z_2$ symmetry
by a remaining global $U(1)$ symmetry
after the $U(1)_\BL$ breaking.
 A Dirac fermion $\Psi_1$ or a complex neutral scalar $s^0_1$
is the dark matter candidate in this model.
 Since the dark matter ($\Psi_1$ or $s^0_1$) has its own $\BL$ charge,
the invisible decay of the $U(1)_\BL$ gauge boson $Z^\prime$
is enhanced.
 Experimental constraints on the model are considered,
and the collider phenomenology
at the LHC as well as future linear colliders
is discussed briefly.
\end{abstract}


\maketitle

\section{Introduction}
\label{sec:intro}

 Neutrino oscillation measurements%
~\cite{solar,Wendell:2010md,acc,Apollonio:2002gd,Gando:2010aa}
have established evidence for tiny neutrino masses,
which are supposed to be zero
in the standard model~(SM) of particle physics.
 It seems mysterious that
the scale of neutrino masses is much smaller than
that of the other fermion masses.
 The simplest way to obtain tiny neutrino masses
is the seesaw mechanism~\cite{seesaw}
where right-handed neutrinos are introduced.
 Due to suppression with
huge Majorana masses of the right-handed neutrinos
as compared to the electroweak scale,
neutrino masses can be very small
even though Dirac Yukawa coupling constants for neutrinos
are of ${\mathcal O}(1)$.
 However
testability of the mechanism seems to be a problem
because key particles (right-handed neutrinos)
with such huge masses
would not be accessible in future experiments.
 A possible solution for the problem
is radiative generation of Dirac Yukawa couplings
for neutrinos.
 In order to explain tiny neutrino masses,
the right-handed neutrinos
with masses of ${\mathcal O}(100)\,\GeV$
are acceptable naturally
without assuming excessive fine tuning
among coupling constants
by virtue of loop-suppressed Dirac Yukawa couplings.
 Radiative generation of Dirac Yukawa couplings
has been discussed in various frameworks such as
the left-right symmetry%
~\cite{Mohapatra:1987hh,1loopLR},
supersymmetry~(SUSY)~\cite{1loopSUSY},
extended models
within the SM gauge group%
~\cite{Nasri:2001ax,Wei:2010ww,Kanemura:2011jj},
and an extra $U(1)$ gauge symmetry~\cite{Gu:2007ug}
(See also ref.~\cite{Babu:1989fg}).

 On the other hand,
existence of dark matter has been indicated
by Zwicky~\cite{Zwicky:1933gu},
and its thermal relic abundance has been
quantitatively determined
by the WMAP experiment~\cite{WMAP}.
 If the essence of the dark matter is an elementary particle,
the weakly interacting massive particle~(WIMP)
would be a promising candidate.
 A naive power counting shows that
the WIMP dark matter mass should be at the electroweak scale.
 This would suggest a strong connection
between the WIMP dark matter and the Higgs sector.
 It is desired to have viable candidate for dark matter
in models beyond the standard model.
 Usually
stability of the dark matter candidate
is ensured by imposing a $Z_2$ parity
where the dark matter is the lightest $Z_2$-odd particle.
 It is well known that
such $Z_2$ odd particles are compatible
with radiative neutrino mass models%
~\cite{Gu:2007ug, KNT, Ma, Ma:2008ba, AKX,Kanemura:2010bq}.
 Usually in such models, however,
the origin of the $Z_2$ parity
has not been clearly discussed.
 It seems better 
if a global symmetry to stabilize the dark matter
is not just imposed additionally
but obtained as a remnant of some broken symmetry
which is used also for other purposes~\cite{Ma:2008ba}.

 In this paper
we propose a new model
in which tiny neutrino masses
and the origin of dark matter
are naturally explained at the TeV-scale.
 We introduce
the $U(1)_\BL$ gauge symmetry
to the SM gauge group
which is spontaneously broken at multi-TeV scale%
~\cite{BL,Kanemura:2011vm,Okada:2010wd}.
 Its collider phenomenology has been studied%
~\cite{Basso:2008iv,Basso:2010pe,Basso:2010si,Basso:2011hn}.
 In our model,
Dirac Yukawa couplings for neutrinos
are forbidden at the tree level by the $U(1)_\BL$.
 They are generated at the one-loop level
after the $U(1)_\BL$ breaking.
 Simultaneously,
right-handed neutrino masses
are generated at the tree level
by spontaneous breaking of the $U(1)_\BL$.
 As a result,
light neutrino masses are obtained effectively
at the two-loop level
without requiring too small coupling constants ($\lesssim 10^{-3}$)
from the TeV-scale physics.
 Furthermore
it turns out that
the dark matter in our model is stabilized
by an unbroken global $U(1)$ symmetry
which appears automatically in the Lagrangian
with appropriate assignments of the $U(1)_\BL$-charges
for new particles.
 The mass of a dark matter candidate $\Psi_1$,
which is a Dirac fermion,
is also generated by the $U(1)_\BL$ breaking~
(See also ref.~\cite{Lindner:2011it}%
\footnote
{
 The dark matter in ref.~\cite{Lindner:2011it}
does not contribute to
the mechanism to generate light neutrino masses.
}).
 We show that the model is viable
under the current experimental constraints.
 Prospects in collider experiments are also discussed.

\section{The model}
\label{sec:model}

\begin{table}[t]
\begin{center}
\begin{tabular}
{|p{25mm}|@{\vrule width 1.8pt\ }p{15mm}|p{15mm}|p{15mm}|p{15mm}|p{15mm}|p{15mm}|}
   \hline
    Particles
     & $s^0$
     & $\eta$
     & $(\Psi_{R})_i$
     & $(\Psi_{L})_i$
     & $(\nu_R^{})_i$
     & $\sigma^0$
    \\ \noalign{\hrule height 1.8pt}
    SU(3)$_{\rm C}$
     & {\bf \underline{1}}
     & {\bf \underline{1}}
     & {\bf \underline{1}}
     & {\bf \underline{1}}
     & {\bf \underline{1}}
     & {\bf \underline{1}}
    \\ \hline
    SU(2)$_{\rm L}$
     & {\bf \underline{1}}
     & {\bf \underline{2}}
     & {\bf \underline{1}}
     & {\bf \underline{1}}
     & {\bf \underline{1}}
     & {\bf \underline{1}}
    \\ \hline
    U(1)$_{\rm Y}$
     & 0
     & $1/2$
     & 0
     & 0
     & 0
     & 0 
    \\ \hline
    U(1)$_{\rm B-L}$
     & $1/2$
     & $1/2$
     & $-1/2$
     & $3/2$
     & $1$
     & $2$
    \\ \hline
\end{tabular}
\caption{
 New particles
and their properties under gauge symmetries of the model.
}
\label{table:particle}
\end{center}
\end{table}

 In our model,
the $U(1)_\BL$ gauge symmetry is added
to the SM gauge group.
 New particles
and their properties under gauge symmetries of the model 
are shown in Table~\ref{table:particle}.
 Fields $s^0$, $\eta$, and $\sigma^0$ are complex scalars
while $(\Psi_R)_i$, $(\Psi_L)_i$, and $(\nu_R^{})_i\ (i=1,2)$
are Weyl fermions.
 All of them except $\eta$~[$= (\eta^+, \eta^0)^T$]
are singlet fields under the SM gauge group.
 The SM Higgs doublet field
$\Phi$~[$= (\phi^+, \phi^0)^T$]
and $\eta$ have different $U(1)_\BL$-charges
although their representations for the SM gauge group are the same.
 Notice that mass terms of $\Psi_R$, $\Psi_L$, and $\nu_R^{}$
are forbidden by the $U(1)_\BL$ symmetry.

 Yukawa interactions are given by
\begin{eqnarray}
 {\cal L}_{\text{Yukawa}}
 &=&
 {\cal L}_{\text{SM-Yukawa}}
 -
 (y_R^{})_i\,
 \overline{ (\nu_R^{})^c_i }\, (\nu_R^{})_i\, (\sigma^0)^\ast
 -
 (y_{\Psi}^{})_i\,
 \overline{ (\Psi_R)_i }\, (\Psi_L)_i\, (\sigma^0)^\ast
 \nonumber\\
 &&\hspace*{-2mm}
{}-
 (y_3)_{ij}\,
 \overline{ (\nu_R^{})^c_i }\, (\Psi_R)_j\, (s^0)^\ast
 -
 h_{ij}\,
 \overline{ (\Psi_L)_i }\, (\nu_R^{})_j\, s^0
 -
 f_{\ell i}\,
 \overline{ (L_L)_\ell }\, (\Psi_R)_i\, i\sigma_2\, \eta^\ast
 + \text{h.c.} ,
 \label{eq:Lagrangian}
\end{eqnarray}
where ${\cal L}_{\text{SM-Yukawa}}$ stands for
the Yukawa interactions in the SM
and $(L_L)_\ell$ are the lepton doublet fields
of flavor $\ell\ (\ell= e,\mu,\tau)$.
 Superscript $c$ means the charge conjugation
and $\sigma_i$~($i=1\text{-}3$) are the Pauli matrices.
 We take a basis
where Yukawa matrices $y_R^{}$ and $y_\Psi^{}$ are diagonalized
such that their real positive eigenvalues satisfy
$(y_R^{})_1 \leq (y_R^{})_2$ and $(y_\Psi^{})_1 \leq (y_\Psi^{})_2$.

 Scalar potential in this model is expressed as
\begin{eqnarray}
 V(\Phi,s.\eta,\sigma)
 &=&
 -\mu_{\phi}^2\Phi^{\dagger} \Phi
 +
 \mu_s^2 |s^0|^2
 +
 \mu_{\eta}^2\eta^{\dagger} \eta
 -
 \mu_{\sigma}^2 |\sigma^0|^2
 \nonumber\\
 &&{}
 +
 \lambda_\phi \left(\Phi^{\dagger} \Phi\right)^2
 +
 \lambda_s |s^0|^4
 +
 \lambda_\eta \left(\eta^{\dagger} \eta\right)^2
 +
 \lambda_\sigma |\sigma^0|^4
 \nonumber\\
 &&{}
 +
 \lambda_{s\eta} |s^0|^2 \eta^{\dagger} \eta
 +
 \lambda_{s\phi} |s^0|^2 \Phi^{\dagger} \Phi
 +
 \lambda_{\phi\phi} (\eta^{\dagger}\eta) (\Phi^{\dagger} \Phi)
 +
 \lambda_{\eta\phi} (\eta^{\dagger} \Phi) (\Phi^{\dagger}\eta)
 \nonumber\\
 &&{}
 +
 \lambda_{s\sigma} |s^0|^2 |\sigma^0|^2
 +
 \lambda_{\sigma\eta} |\sigma^0|^2 \eta^{\dagger} \eta
 +
 \lambda_{\sigma\phi} |\sigma^0|^2 \Phi^{\dagger} \Phi
 +\left( \mu_3^{}\, s^0\, \eta^{\dagger}\, \Phi + \text{h.c.}\right) ,
 \label{eq:V}
 \end{eqnarray}
where $\mu_\phi^2$, $\mu_\sigma^2$, $\mu_s^2$, and $\mu_\eta^2$
are positive values.
 The coupling constant $\mu_3^{}$ of the trilinear term
can be taken as a real positive value
by redefinition of phase of $s^0$.
 The $\mu_3^{}$ is the breaking parameter
for a global $U(1)_{\eta-\text{L}}$ which conserves
difference between the $\eta$ number and the SM lepton number.
 Some coupling constants in the potential
are constrained by the tree-level unitarity~\cite{Basso:2010jt}.
 Notice that
this model has a global $U(1)$ symmetry (we refer to it as $U(1)_\DM$)
of which $s^0$, $\eta$, $\Psi_R$, and $\Psi_L$ have the same charge
and others have no charge.
 Because of the global $U(1)_\DM$ symmetry,
the Lagrangian is not changed
by an overall shift of $U(1)_\BL$-charges
with an integer for the $U(1)_\DM$-charged particles.

 The $U(1)_\BL$ is broken spontaneously
by the vacuum expectation value (vev) of $\sigma^0$,
$v_\sigma$~[$= \sqrt{2} \av{\sigma^0}$]
while
the $SU(2)_L\times U(1)_Y$ is broken to $U(1)_\EM$
by the vev of $\phi^0$,
$v$~[$=\sqrt{2} \av{\phi^0} \simeq 246\,\GeV$].  
 By imposing the stationary condition,
$v_\sigma$ and $v$ are determined as
\begin{eqnarray}
\begin{pmatrix}
 v^2\\
 v_\sigma^2  
\end{pmatrix}
=
 \frac{1}{\lambda_\sigma \lambda_\phi - \lambda_{\sigma\phi}^2/4}
 \begin{pmatrix}
  \lambda_\sigma & -\lambda_{\sigma\phi}/2\\
  -\lambda_{\sigma\phi}/2 & \lambda_\phi
 \end{pmatrix}
 \begin{pmatrix}
  \mu_\phi^2\\
  \mu_\sigma^2
 \end{pmatrix} .
\end{eqnarray}
 The gauge boson $Z^\prime$ of $U(1)_\BL$
acquires its mass as $m_{Z^\prime} = 2 g_\BL^{} v_\sigma$,
where $g_\BL^{}$ denotes gauge coupling constant of $U(1)_\BL$.
 A constraint $v_\sigma > 3.5\,\TeV$ is given 
by precision tests of the electroweak interaction~\cite{LEPZp}.
 Furthermore,
right-handed neutrinos $(\nu_R^{})_i$ obtain Majorana masses
$(m_R^{})_i$~[$= \sqrt{2} (y_R^{})_i v_\sigma$] 
while $(\Psi_R)_i$ and $(\Psi_L)_i$ for each $i$
become a Dirac fermion $\Psi_i$
with its mass $m_{\Psi_i}^{}$~[$= (y_\Psi^{})_i v_\sigma/\sqrt{2}$].
 Since the global $U(1)_\DM$ is not broken by $v_\sigma$,
the lightest $U(1)_\DM$-charged particle is stable.
 Notice that
there is no anomaly for the $U(1)_\DM$
because $(\Psi_R)_i$ and $(\Psi_L)_i$ have the same $U(1)_\DM$-charge.
 If the particle is electrically neutral
($\Psi_1$ or a mixture of $s^0$ and $\eta^0$),
it becomes a candidate for the dark matter.

 After symmetry breaking with $v_\sigma$ and $v$,
mass eigenstates of two CP-even scalars
and their mixing angle $\alpha$
are given by
\begin{eqnarray}
\begin{pmatrix}
 h^0\\
 H^0
\end{pmatrix}
=
 \begin{pmatrix}
  \cos\alpha & -\sin\alpha\\
  \sin\alpha & \cos\alpha
 \end{pmatrix}
 \begin{pmatrix}
  \phi^0_r\\
  \sigma^0_r
 \end{pmatrix} , \quad
%
\sin{2\alpha}
=
 \frac{ 2\lambda_{\sigma\phi} v v_\sigma }{ m_{H^0}^2 - m_{h^0}^2 } ,
\end{eqnarray}
where $\sigma^0 = (v_\sigma + \sigma^0_r + iz_\sigma)/\sqrt{2}$
and $\phi^0 = (v + \phi^0_r + iz_\phi)/\sqrt{2}$.
 It is needless to say that
$z_\phi$ and $z_\sigma$ are Nambu-Goldstone bosons
which are absorbed by $Z$ and $Z^\prime$, respectively.
 Masses of $h^0$ and $H^0$ are defined by
\begin{eqnarray}
m_{h^0}^2
&=&
 \lambda_\phi v^2 + \lambda_\sigma v_\sigma^2
 -\sqrt{
   \left(
    \lambda_\phi v^2 - \lambda_\sigma v_\sigma^2
   \right)^2
   + \lambda_{\sigma\phi}^2 v^2 v_\sigma^2 }\, ,
 \nonumber\\
%
m_{H^0}^2
&=&
 \lambda_\phi v^2 + \lambda_\sigma v_\sigma^2
 +\sqrt{
   \left(
    \lambda_\phi v^2 - \lambda_\sigma v_\sigma^2
   \right)^2
   + \lambda_{\sigma\phi}^2 v^2 v_\sigma^2 }\, .
\end{eqnarray}
 On the other hand,
since $s^0$ and $\eta^0$ are $U(1)_\DM$-charged particles,
they are not mixed with $\sigma^0$ and $\phi^0$. 
 Mass eigenstates of these $U(1)_\DM$-charged scalars
and their mixing angle $\theta$
are obtained as
\begin{eqnarray}
\begin{pmatrix}
 s^0_1\\
 s^0_2
\end{pmatrix}
=
 \begin{pmatrix}
  \cos\theta & -\sin\theta\\
  \sin\theta & \cos\theta
 \end{pmatrix}
 \begin{pmatrix}
  \eta^0\\
  s^0
 \end{pmatrix} , \quad
%
\sin{2\theta}
=
 \frac{ \sqrt{2} \mu_3^{} v }{ m_{s^0_2}^2 - m_{s^0_1}^2 } .
\end{eqnarray}
 Mass eigenvalues $m_{s^0_1}$ and $m_{s^0_2}$
of these neutral complex scalars
are defined by
\begin{eqnarray}
m_{s^0_1}^2
&=&
 \frac{1}{2}
 \left(
  m_\eta^2 + m_s^2
  -\sqrt{ \left( m_\eta^2 - m_s^2 \right)^2 + 2 \mu_3^2 v^2 }
 \right),
\nonumber\\
%
m_{s^0_2}^2
&=&
 \frac{1}{2}
 \left(
  m_\eta^2 + m_s^2
  +\sqrt{ \left( m_\eta^2 - m_s^2 \right)^2 + 2 \mu_3^2 v^2 }
 \right) ,
\end{eqnarray}
where
$m_s^2
= \mu_s^2 + \lambda_{s\phi} v_{\phi}^2/2 + \lambda_{s\sigma} v_{\sigma}^2/2$
and
$m_{\eta}^2
= \mu_{\eta}^2
 + \left( \lambda_{\phi\phi} + \lambda_{\eta\phi} \right) v_{\phi}^2/2
 + \lambda_{\sigma\eta} v_{\sigma}^2/2$.
 Finally,
the mass of charged scalars $\eta^\pm$ is
\begin{eqnarray}
m_{\eta^{\pm}}^2
&=&
 \mu_\eta^2 +\lambda_{\phi\phi} \frac{v^2}{\,2\,}
 + \lambda_{\sigma\eta} \frac{v_{\sigma}^2}{2} .
 \label{eq:m_etapm}
\end{eqnarray}

\section{Neutrino Mass and Dark Matter}
\label{sec:nuDM}

\subsection{Neutrino Mass}
\label{subsec:numass}

\begin{figure}[t]
 \begin{center}
  \scalebox{0.28}{\includegraphics{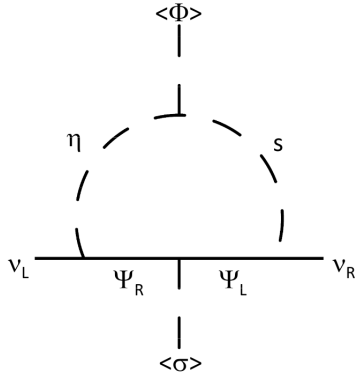}}
\vspace*{-7mm}
  \caption{Diagram for Dirac mass terms of neutrinos.}
  \label{fig:1-loop}
 \end{center}
\end{figure}

\begin{figure}[t]
 \begin{center}
  \scalebox{0.32}{\includegraphics{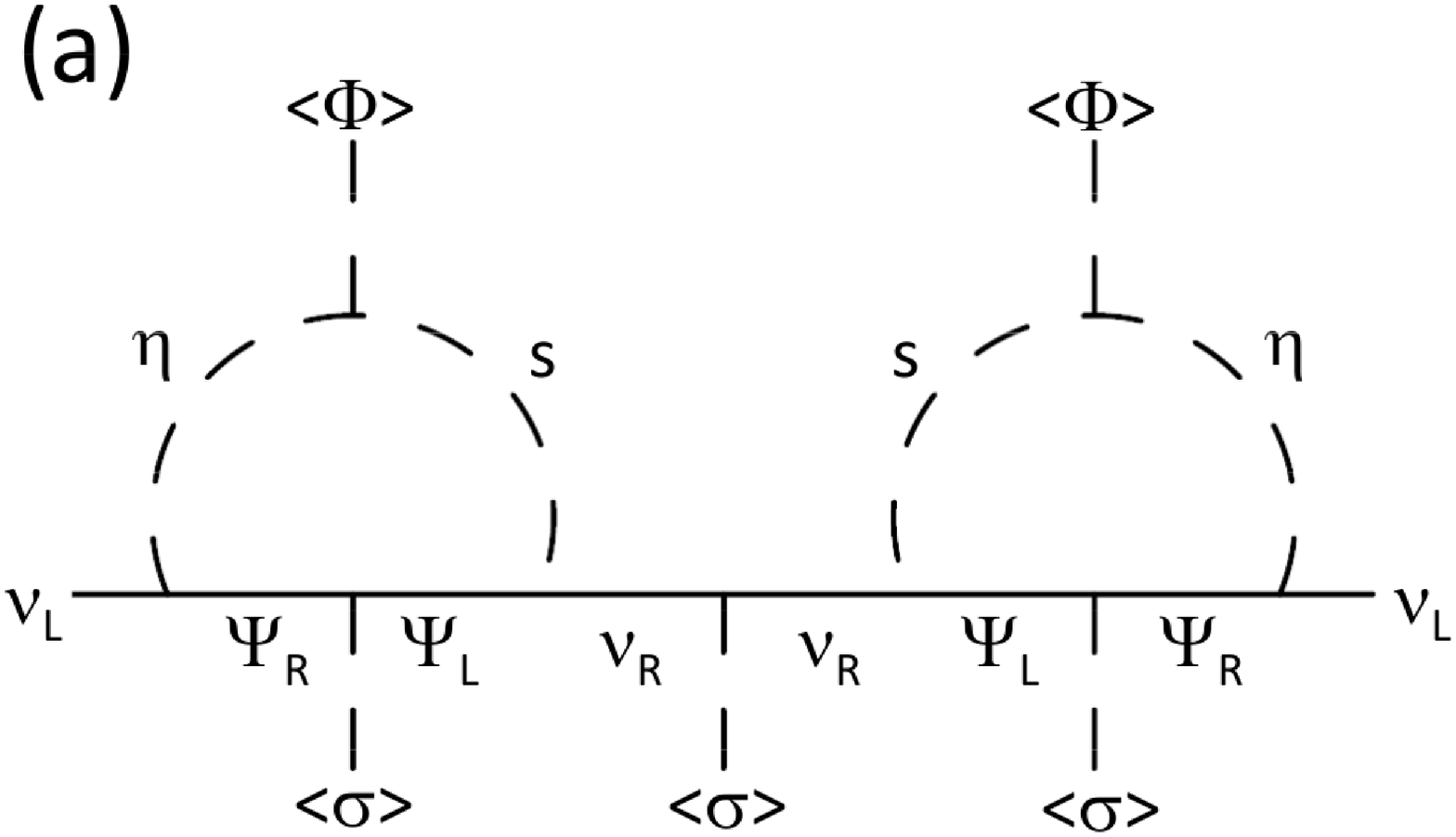}}\\[-10mm]
  \scalebox{0.28}{\includegraphics{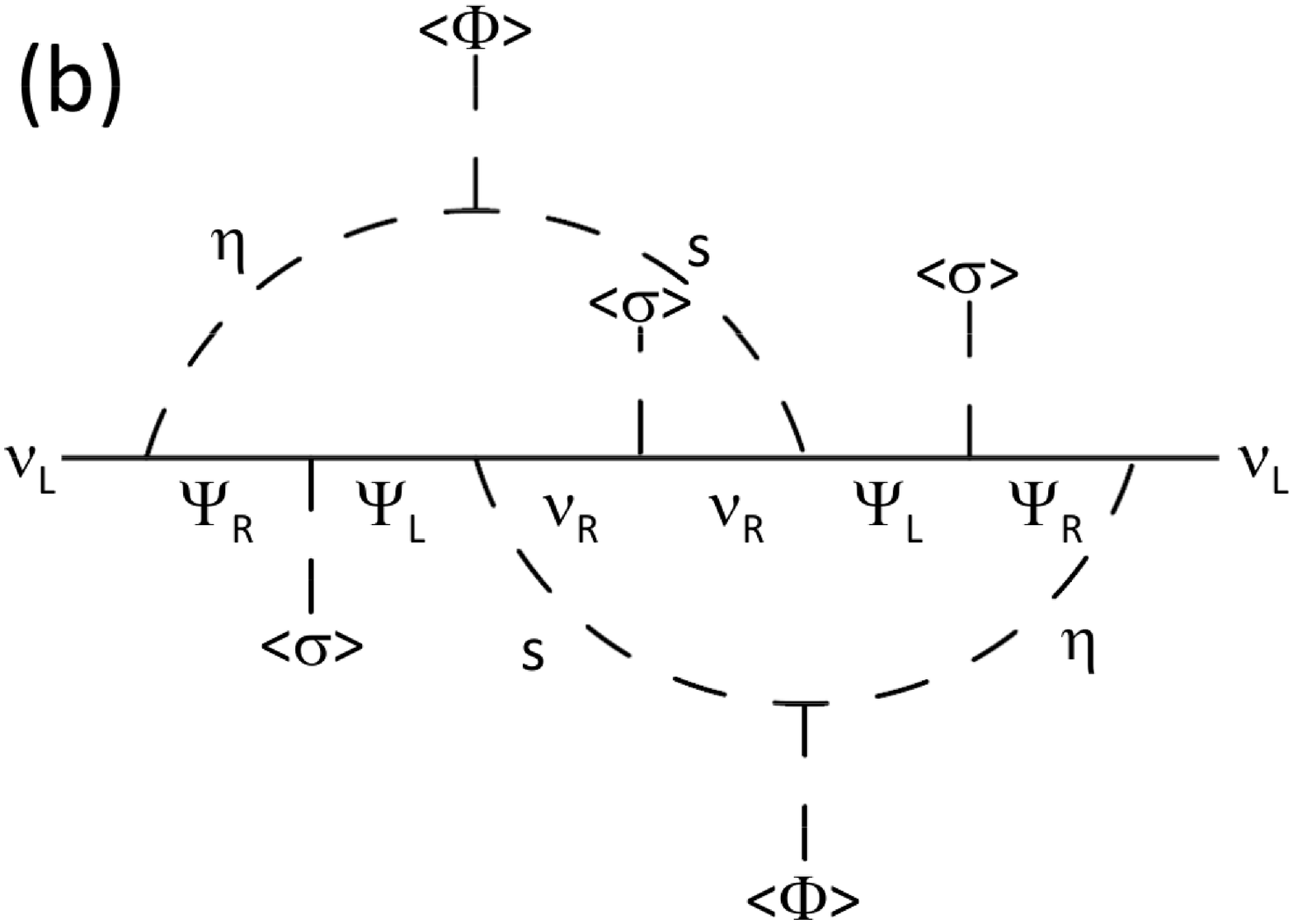}}\hspace*{0mm}
  \scalebox{0.28}{\includegraphics{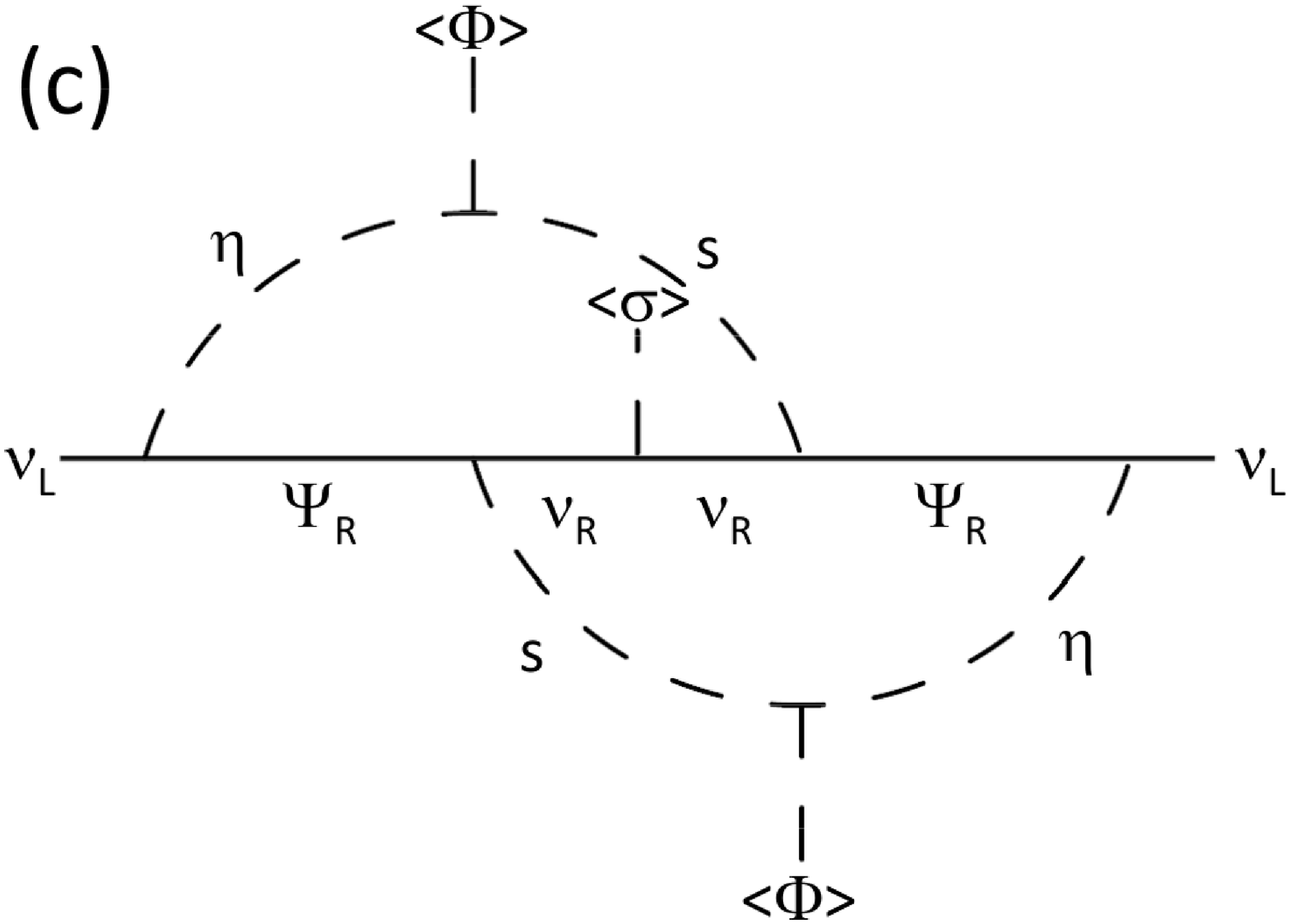}}
\vspace*{-5mm}
  \caption{Diagrams for light Majorana neutrino masses.}
  \label{fig:neutrinomass}
 \end{center}
\end{figure}

 In this model,
Dirac mass terms for neutrinos
are generated by a one-loop diagram
in Fig.~\ref{fig:1-loop}.
 This diagram is used also
in a model in ref.~\cite{Gu:2007ug}
in which lepton number is conserved.
 Via the seesaw mechanism,
tiny Majorana masses of light neutrinos
are induced at the two-loop level
as shown in Fig.~\ref{fig:neutrinomass}(a)
(See also refs.~\cite{Nasri:2001ax,Wei:2010ww}).
 In addition,
there are one-particle-irreducible~(1PI) diagrams also
at the two-loop level%
~(Figs.~\ref{fig:neutrinomass}(b) and \ref{fig:neutrinomass}(c))%
\footnote{
 Such 1PI diagrams were overlooked in refs.~\cite{Nasri:2001ax,Wei:2010ww}.
}.
 The Majorana mass matrix
is calculated as
\begin{eqnarray}
\left( m_{\nu} \right)_{\ell\pell}
&=&
 \left(
  \frac{1}{16\pi^2}
 \right)^2
 \Biggl\{
  \sum_{i,j,a}
  f_{\ell i}\,
  h_{ia}\, (m_R^{})_a\, (h^T)_{aj}\,
  (f^T)_{j\pell}
  \Bigl[
   \left( I_1 \right)_{ija}
   + \left\{ I_2 \right\}_{ija}
  \Bigr]
\nonumber\\
&&\hspace*{40mm}
{}+
 \sum_{i,j,a}
 f_{\ell i}\,
 (y_3^\dagger)_{ia}\, (m_R^{})_a\, (y_3^\ast)_{aj}\,
 (f^T)_{j\pell}
 \left\{ I_3 \right\}_{ija}
 \Biggr\} ,
 \label{eq:neutrinomass}
\end{eqnarray}
where
dimensionless functions $I_1$, $I_2$, and $I_3$
are defined by
\begin{eqnarray}
(I_1)_{ija}
&=&
 -
 \frac{ ( 8\pi^2 \sin2\theta )^2 m_{\Psi_i}^{} m_{\Psi_j}^{} }
      { (m_R^{})_a^2 }
 \left[
  \int\!\!\frac{d^4k_1}{(2\pi)^4}
  \frac{1}{ k_1^2 - m_{\Psi_i}^2 }
  \left\{
   \frac{1}{ k_1^2 - m_{s^0_1}^2 }
   -
   \frac{1}{ k_1^2 - m_{s^0_2}^2 }
  \right\}
 \right]
\nonumber\\
&&\hspace*{40mm}
 \times
 \left[
  \int\frac{d^4k_2}{(2\pi)^4}
  \frac{1}{ k_2^2 - m_{\Psi_j}^2 }
  \left\{
   \frac{1}{ k_2^2 - m_{s^0_1}^2 }
   -
   \frac{1}{ k_2^2 - m_{s^0_2}^2 }
  \right\}
 \right] ,
\\
(I_2)_{ija}
&=&
 ( 8\pi^2 \sin2\theta )^2 m_{\Psi_i}^{} m_{\Psi_j}^{}
 \nonumber\\
&&\times
 \int\!\!\!\int\frac{d^4k_1}{(2\pi)^4}\frac{d^4k_2}{(2\pi)^4}
 \left\{
  \frac{1}{ k_1^2 - m_{s^0_1}^2 }
  -
  \frac{1}{ k_1^2 - m_{s^0_2}^2 }
 \right\}
 \frac{1}{k_1^2-m_{\Psi_i}^2}
 \nonumber\\
&&\times
 \frac{1}{(k_1-k_2)^2-(m_R^{})_a^2}
 \left\{
  \frac{1}{ k_2^2 - m_{s^0_1}^2 }
  -
  \frac{1}{ k_2^2 - m_{s^0_2}^2 }
 \right\}
 \frac{1}{k_2^2 - m_{\Psi_j}^2} ,
\\
(I_3)_{ija}
&=&
 ( 8\pi^2 \sin2\theta )^2
 \int\!\!\!\int\frac{d^4k_1}{(2\pi)^4}\frac{d^4k_2}{(2\pi)^4}
 k_1\cdot k_2
 \left\{
  \frac{1}{ k_1^2 - m_{s^0_1}^2 }
  -
  \frac{1}{ k_1^2 - m_{s^0_2}^2 }
 \right\}
 \frac{1}{k_1^2-m_{\Psi_i}^2}
 \nonumber\\
&&\times
 \frac{1}{(k_1-k_2)^2-(m_R^{})_a^2}
 \left\{
  \frac{1}{ k_2^2 - m_{s^0_1}^2 }
  -
  \frac{1}{ k_2^2 - m_{s^0_2}^2 }
 \right\}
 \frac{1}{k_2^2 - m_{\Psi_j}^2} ,
\end{eqnarray}
which correspond to the diagrams
in Figs.~\ref{fig:neutrinomass}(a), \ref{fig:neutrinomass}(b),
and \ref{fig:neutrinomass}(c),
respectively.

 If there is only one $\Psi$ (one $\Psi_L$ and one $\Psi_R$),
the mass matrix $(m_\nu)_{\ell\pell}$
is proportional to $f_\ell f_\pell$.
 Then
two of three eigenvalues of $(m_\nu)_{\ell\pell}$ are zero
and the mass matrix conflicts with the oscillation data.
 Therefore
two $\Psi_i$ are introduced in this model.
 We also introduce two $(\nu_R^{})_i$
in order for an easy search of parameter sets
which satisfy experimental constraints%
\footnote{
 With two $\Psi_i$,
one $\nu_R^{}$ is sufficient for that
the rank of $(m_\nu)_{\ell\pell}$ is two.
}.
 Then
the rank of $(m_\nu)_{\ell\pell}$ is two,
for which one neutrino becomes massless.
 Hereafter
degeneracy of right-handed neutrino masses,
$m_R^{} \equiv (m_R^{})_1 = (m_R^{})_2$,
is assumed for simplicity.

 The mass matrix $(m_\nu)_{\ell\pell}$
is diagonalized by the Maki-Nakagawa-Sakata~(MNS) matrix $U_\MNS$
as $U_\MNS^T\, m_\nu\, U_\MNS = \text{diag}(m_1, m_2, m_3)$.
 The standard parametrization of the MNS matrix is
\begin{eqnarray}
U_\MNS
=
 \begin{pmatrix}
  1 & 0 & 0\\
  0 & c_{23} & s_{23}\\
  0 & -s_{23} & c_{23} 
 \end{pmatrix}
 \begin{pmatrix}
  c_{13} & 0 & s_{13}\, e^{-i\delta}\\
  0 & 1 & 0\\
  -s_{13}\, e^{i\delta} & 0 & c_{13} 
 \end{pmatrix}
 \begin{pmatrix}
  c_{12} & s_{12} & 0\\
  -s_{12} & c_{12} & 0\\
  0 & 0& 1 
 \end{pmatrix} ,
\label{eq:MNS}
\end{eqnarray}
where $c_{ij}$ and $s_{ij}$ stand for
$\cos{\theta_{ij}}$ and $\sin{\theta_{ij}}$,
respectively.
 Mixing angles $\theta_{ij}$
and $\Delta m^2_{ij} \equiv m_i^2 - m_j^2$
are constrained by neutrino oscillation measurements%
~\cite{solar,Wendell:2010md,acc,Apollonio:2002gd,Gando:2010aa}.
 In our analyses
we use the following values as an example;
\begin{eqnarray}
&&
s_{23}^2 = \frac{1}{\,2\,} , \quad
s_{13}^2 = 0 , \quad
s_{12}^2 = \frac{1}{\,3\,} ,
\label{eq:mix}\\
&&
\Delta m^2_{21} = 7.5\times 10^{-5}\,\eV^2 , \quad
|\Delta m^2_{31}| = 2.3\times 10^{-3}\,\eV^2 .
\label{eq:Dmsq}
\end{eqnarray}
 Notice that there is no difficulty
to use nonzero values of $s_{13}$~\cite{Abe:2011sj}
in our analyses.
 In Table~\ref{table:parameter},
we show two examples~(Set~A and Set~B) for the parameter set
which reproduces the values given
in eqs.~\eqref{eq:mix} and \eqref{eq:Dmsq}
for $\Delta m^2_{31} > 0$.
 These sets satisfy also other experimental constraints
as shown below.

\begin{table}[t]
\begin{center}
\begin{tabular}{|c|@{\vrule width 1.8pt}c|c|}
\hline
 {} & Set A & Set B\\
\noalign{\hrule height 1.8pt}
 $f_{\ell i}$
  &\rule[0mm]{0mm}{15mm}
   $
    \begin{pmatrix}
     -0.00726 \ & 0.00667\\
     -0.0523 \ & 0.0206\\
     -0.0378 \ & 0.00723
    \end{pmatrix}
   $
  &
   $
    \begin{pmatrix}
     -0.0485 \ & 0.0505\\
     -0.0364 \ & 0.0433\\
     0.0606 \ & -0.0577
    \end{pmatrix}
   $\\[10mm]
\hline
 $h_{ij}$
  &\rule[0mm]{0mm}{10mm}
   $
    \begin{pmatrix}
     -0.119 \ & 0.150\\
     0.150 \ & 0.150\\
    \end{pmatrix}
   $
  &
   $
    \begin{pmatrix}
     0.544 \ & 0.505\\
     0.505 \ & 0.505\\
    \end{pmatrix}
   $\\[5mm]
\hline
 $(y_3)_{ij}$
  &\rule[0mm]{0mm}{10mm}
   $
    \begin{pmatrix}
     0.0152 \ & 0.0152\\
     0.0152 \ & 0.0152\\
    \end{pmatrix}
   $
  &
   $
    \begin{pmatrix}
     0.0101 \ & 0.0101\\
     0.0101 \ & 0.0101\\
    \end{pmatrix}
   $\\[5mm]
\hline
 \ $m_R^{} \equiv (m_R^{})_1 = (m_R^{})_2$ \
  & $250\,\GeV$
  & $200\,\GeV$\\
\hline
 $\{ m_{\Psi_1}^{},\ m_{\Psi_2}^{} \}$
  & $\{ 57.0\,\GeV,\ 800\,\GeV \}$
  & $\{ 800\,\GeV,\ 800\,\GeV \}$\\
\hline
 $\{ m_{h^0}^{},\ m_{H^0}^{},\ \cos\alpha \}$
  & \ $\{ 120\,\GeV,\ 140\,\GeV,\ 1/\sqrt{2} \}$ \
  & $\{ 130\,\GeV,\ 300\,\GeV,\ 1 \}$\\
\hline
 $\{ m_{s^0_1},\ m_{s^0_2},\ \cos\theta \}$
  & $\{ 200\,\GeV,\ 300\,\GeV,\ 0.05 \}$
  & \ $\{ 55.0\,\GeV,\ 250\,\GeV,\ 0.05 \}$ \ \\
\hline
 $m_{\eta^\pm}$
  & $280\,\GeV$
  & $220\,\GeV$\\
\hline
 $g_\BL^{}$
  & $0.2$
  & $0.2$\\
\hline
 $m_{Z^\prime}^{}$
  & $2000\,\GeV$
  & $2000\,\GeV$\\
\hline
\end{tabular}
\end{center}
\caption{
 Two examples of the parameter set
which satisfies experimental constraints.
 The dark matter is $\Psi_1$ for Set~A
while it is $s^0_1$~($\simeq s^0$) for Set~B\@.
}
\label{table:parameter}
\end{table}

\subsection{Lepton Flavor Violation}
\label{subsec:LFV}

 Charged scalar bosons $\eta^\pm$,
which also have a $U(1)_\DM$-charge,
contribute to the $\meg$ process.
 The branching ratio of $\meg$ is calculated as
\begin{eqnarray}
\BR(\mu\to e \gamma)
&=&
 \frac{3\alpha_\EM^{}}{64\pi G_F^2}
 \left|
  \frac{1}{m_{\eta^{\pm}}^2} f_{\mu i}
  F_2\left(
      \frac{ m_{\Psi_i}^2 }{ m_{\eta^{\pm}}^2 }
     \right) (f^\dagger)_{ie}
 \right|^2 ,
 \label{eq:meg}
\end{eqnarray}
where
\begin{eqnarray}
F_2\left(a\right)
&\equiv&
 \frac{ 1 - 6a + 3a^2 + 2a^3 - 6a^2 \ln(a) }
      { 6\left( 1-a \right)^4 }.
\end{eqnarray}
 For Set~A and Set~B,
we obtain
$\BR(\meg) = 5.1\times 10^{-13}$
and $1.7\times 10^{-12}$, respectively.
 They satisfy
the current upper bound;
$\BR(\meg) < 2.4\times 10^{-12}$~(90\% CL)~\cite{Adam:2011ch}.
 These values for Set~A and Set~B
could be within the future experimental reach.

\subsection{Dark Matter}
\label{subsec:DM}

 The dark matter candidate is
$\Psi_1$ (for Set~A)
or $s^0_1$ (for Set~B).
 The relic abundance of dark matter
is constrained stringently by the WMAP experiment
as $\Omega_\DM h^2 \simeq 0.11$~\cite{WMAP}.
 The dark matter candidate in this model
pair-annihilates into a pair of SM fermions $f$
by $s$-channel processes mediated by $h^0$ and $H^0$
for both of Set~A and Set~B\@.
 The $t$-channel diagram is highly suppressed
due to the small values of $f_{\ell i}$,
which are required by the $\meg$ search results.

\begin{figure}[t]
\begin{center}
\includegraphics[angle=-90,scale=0.32]{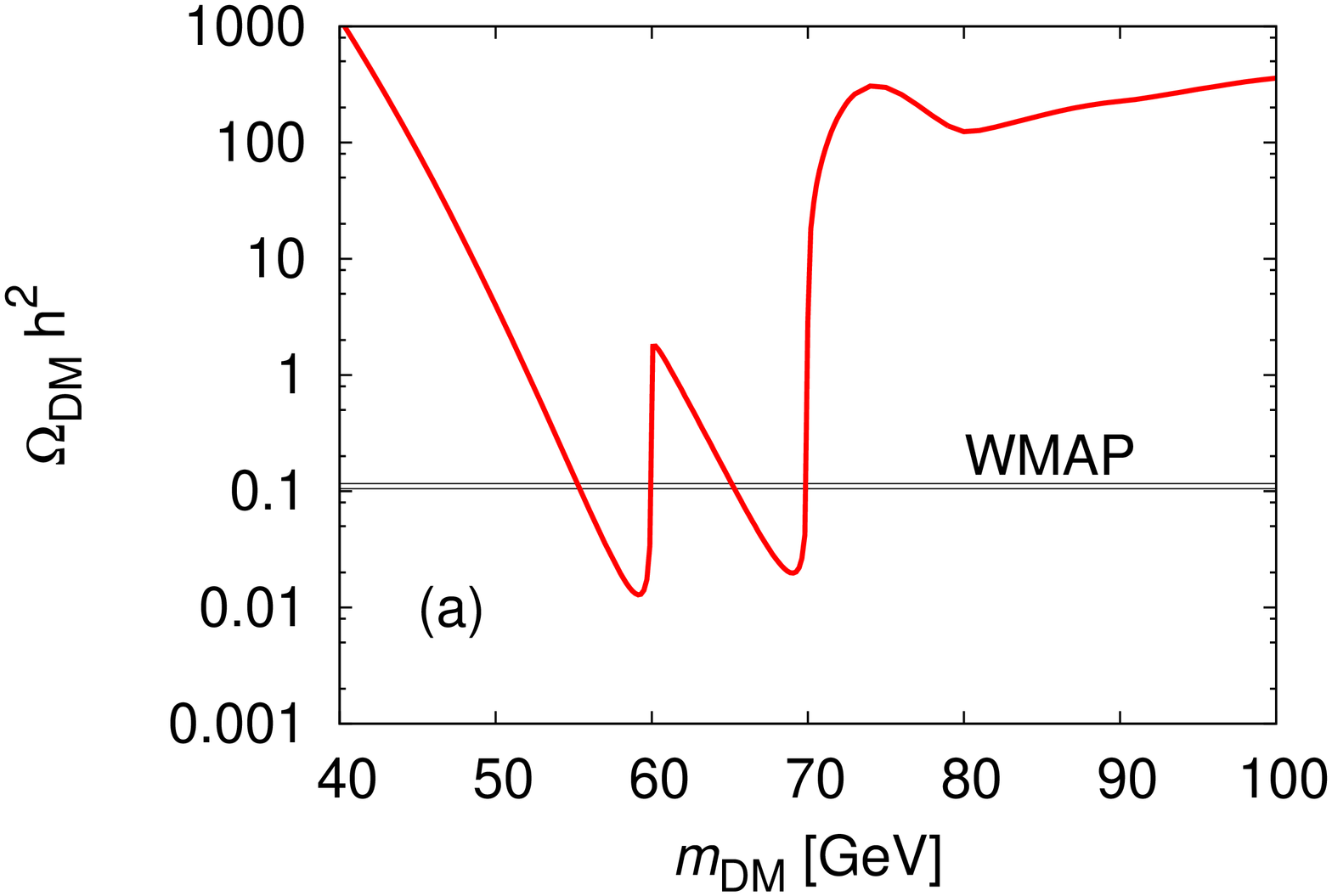}
\includegraphics[angle=-90,scale=0.32]{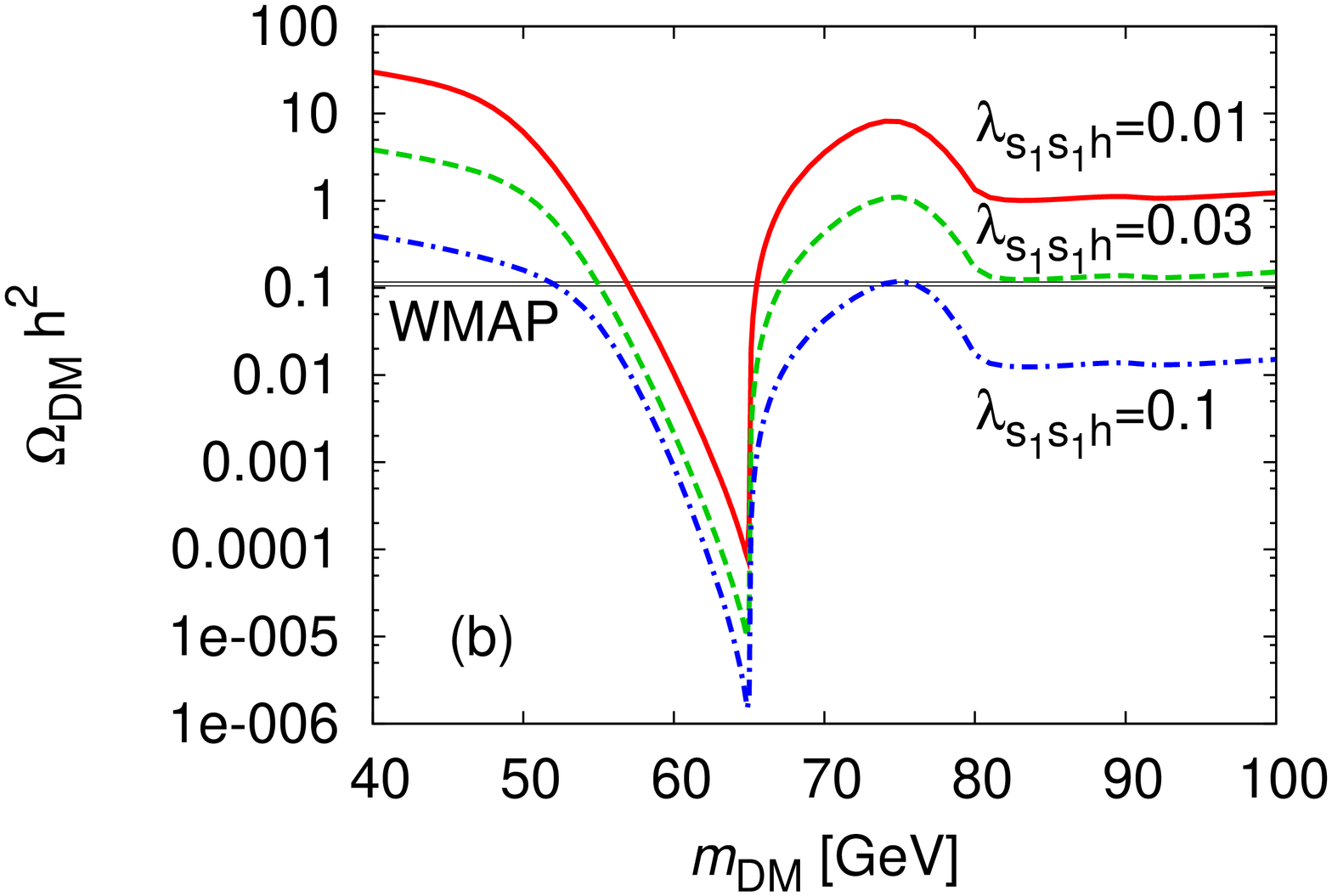}
\caption{
 The relic abundance with respect to
the dark matter mass $m_\DM$.
 Horizontal lines show
$1\sigma$ allowed region
($0.1053 \leq \Omega_\DM h^2 \leq 0.1165$)
of the WMAP result.
 (a) Case for Set~A
where $\Psi_1$ is the dark matter candidate.
 (b) Case for Set~B
where $s^0_1$ is the dark matter candidate.
 Three curves are obtained for
$\lambda_{s_1 s_1 h}^{} =0.1$~(lower curve),
$0.03$~(middle curve), and $0.01$~(upper curve).
}
\label{fig:WMAP}
\end{center}
\end{figure}

 We first consider the case where
$\Psi_1$ is the dark matter, i.e.\ Set~A,
whose mass is given by $(y_{\Psi}^{})_1 v_\sigma/\sqrt{2}$.
 Because $v_\sigma > 3.5\,\TeV$,
the magnitude of $(y_\Psi^{})_1$ is $\lesssim 0.01$
for $m_{\Psi_1}^{} = {\mathcal O}(10)\,\GeV$.
 The annihilation cross section is
suppressed by the small $(y_\Psi^{})_1$
because it is proportional to
$( m_f (y_\Psi^{})_1 \sin{2\alpha}/v )^2$.
 In order to enhance the cross section for the appropriate relic abundance,
a large mixing between $\sigma^0_r$ (which couples with $\Psi_1$)
and $\phi^0_r$ (which couples with SM fermion $f$)
is required~\cite{Okada:2010wd,Kanemura:2011vm}.
 Thus
we take $\cos\alpha = 1/\sqrt{2}$ for Set~A\@.
 Then
the WMAP result gives
a constraint on the dark matter mass $m_{\Psi_1}^{}$
for fixed $m_{h^0}^{}$ and $m_{H^0}^{}$
($120\,\GeV$ and $140\,\GeV$ for Set~A, respectively).
 Figure~\ref{fig:WMAP}(a) shows
dependence of the relic abundance on $m_{\Psi_1}^{}$
where other parameters are the same as values of Set~A\@.
 It can be confirmed in the figure
that our choice $m_{\Psi_1}^{} = 57\,\GeV$
is consistent with the WMAP result for Set~A\@.

 Next,
we consider the case
where $s^0_1$ is the dark matter. 
 A coupling constant $\lambda_{s_1 s_1 h}^{}$
for $\lambda_{s_1 s_1 h}^{}\, v\, s^0_1 (s^0_1)^\ast h^0$
is constrained by the WMAP result.
 Figure~\ref{fig:WMAP}(b) shows
the relic abundance of $s^0_1$
as functions of $m_{s^0_1}$
for several values of $\lambda_{s_1 s_1 h}^{}$.
 In the figure,
we used $m_{h^0}^{} = 130\,\GeV$ and $m_{H^0}^{} = 300\,\GeV$
which are the values for Set~B\@.
 Contribution of $H^0$ to the annihilation cross section
is negligible
because $m_{H^0}^{}$ is taken to be
away from $2 m_{s^0_1} = 110\,\GeV$ for simplicity.
 In order to satisfy the WMAP constraint for Set~B,
we find that
\begin{eqnarray}
\lambda_{s_1 s_1 h}^{}
\simeq 0.03 .
\label{eq:WMAP-B}
\end{eqnarray}
 This constraint can be satisfied easily
because $\lambda_{s_1 s_1 h}^{}$ can be taken to be arbitrary
depending on several parameters in the scalar potential.

 Finally,
we discuss the constraint from
direct search experiments for the dark matter.
 If $s^0_1$ is mainly composed of $\eta^0$,
it cannot be a viable candidate for the dark matter
even if it is the lightest $U(1)_\DM$-charged particle.
 This is because
the scattering cross section with a nucleon $N$~($N=p,n$)
becomes too large due to the weak interaction.
 Thus
$s^0_1$ should be dominantly made from singlet $s^0$,
and this is the reason why we take $\cos\theta = 0.05$ for Set~B\@.
 Scattering cross sections
of dark matter candidates ($\Psi_1$ and $s^0_1$)
with a nucleon $N$~($N=p,n$) are given by
\begin{eqnarray}
\sigma(\Psi_1 N \to \Psi_1 N)
&\simeq&
 \frac{ 8 g_\BL^2 m_{\Psi_1}^2 \sin^2\alpha \cos^2\alpha }
      { v^2 m_{Z^\prime}^2 }
 \left(
  \frac{1}{m_{h^0}^2}
  - \frac{1}{m_{H^0}^2}
 \right)^2
 \frac{ m_N^2 m_{\Psi_1}^2 }
      { \pi \left( m_{\Psi_1}^{} + m_N^{} \right)^2 }\,
 f_N^2
\nonumber\\
&&\hspace*{50mm}
+{}
 \left( \frac{g_\BL^{}}{m_{Z^\prime}} \right)^4
 \frac{ m_{\Psi_1}^2 m_N^2 }
      { 4\pi ( m_{\Psi_1}^{} + m_N^{} )^2 } ,
\label{eq:direct-A}
\\
%
\sigma(s^0_1 N \to s^0_1 N)
&\simeq&
 \frac{1}{\,4\,}
 \left\{
  \frac{ \lambda_{s_1 s_1 h} \cos\alpha }{ m_{h^0}^2}
  +
  \frac{ \lambda_{s_1 s_1 H} \sin\alpha }{ m_{H^0}^2}
 \right\}^2
 \frac{ m_N^2 }
      { \pi \left( m_{s^0_1} + m_N^{} \right)^2 }\,
 f_N^2
\nonumber\\
&&\hspace*{50mm}
{}+
 \left( \frac{g_\BL^{}}{m_{Z^\prime}} \right)^4
 \frac{ m_{s^0_1}^2 m_N^2 }
      { 4 \pi ( m_{s^0_1} + m_N^{} )^2 } ,
\label{eq:direct-B}
\\
%
f_N
&\equiv&
 \sum_{q=u,d,s} m_N^{} f_{Tq} + \frac{2}{\,9\,} m_N^{} f_{Tg} ,
 \label{eq:DMNtoDMN}
\end{eqnarray}
where $m_N^{}$ is the mass of the nucleon
and
 we use
$f_{Tu}+f_{Td} = 0.056$, $f_{Ts} = 0$~\cite{lattice},
and $f_{Tg} = 0.944$~\cite{trace-anomaly}.
 Our results for the $Z^\prime$ mediation are consistent
with those in ref.~\cite{Zheng:2010js}.
 Difference between $p$ and $n$ is neglected.
 We have
$\sigma(\Psi_1\, N \to \Psi_1\, N) = 2.7\times 10^{-45}\,\text{[cm$^2$]}$
for $m_{\Psi_1}^{} = 57\,\GeV$ for Set~A\@.
 The value is dominantly given by
the $Z^\prime$ mediation
while scalar mediations give only
$2.4\times 10^{-48}\,\text{[cm$^2$]}$.
 On the other hand,
for $m_{s^0_1} = 55\,\GeV$ for Set~B,
we have
$\sigma(s^0_1\, N \to s^0_1\, N) = 4.4\times 10^{-45}\,\text{[cm$^2$]}$
by taking into account eq.~\eqref{eq:WMAP-B}
as the WMAP constraint.
 Contributions from $Z^\prime$ and $h^0$
are $2.6\times 10^{-45}\,\text{[cm$^2$]}$
and $1.7\times 10^{-45}\,\text{[cm$^2$]}$,
respectively.
 These values of cross sections for two sets are
just below the constraint from
the XENON100 experiment~\cite{Aprile:2011hi}.
 Notice that even if such values are excluded in near future
this model is not ruled out
because the $Z^\prime$ contribution ($\propto v_\sigma^{-4}$)
can be easily suppressed by
a little bit larger $v_\sigma$.

\section{Prospects for Collider Phenomenology and Discussion}
\label{sec:pheno}

\subsection{Collider Phenomenology}
\label{subsec:coll}

 We have seen that
Set~A and Set~B satisfy
experimental constrains in the previous section.
 Let us discuss the collider phenomenology
by using these parameter sets.

 Since $U(1)_\BL$ is dealt with as the origin
of neutrino masses etc.,
$Z^\prime$ is an important particle in this model.
 The $Z^\prime$ can be the mother particle
to produce the $U(1)_\DM$-charged particles and $\nu_R^{}$ in the model
at collider experiments
because they are all $U(1)_\BL$-charged.
 For $m_{Z^\prime}^{} = 2\,\TeV$ and $g_\BL^{} = 0.2$,
the production cross section of $Z^\prime$
at the CERN Large Hadron Collider~(LHC) with $\sqrt{s}=14\,\TeV$
is $70\,\text{fb}$~\cite{Basso:2008iv,Basso:2011hn}.
 The number of $Z^\prime$ produced with $100\,\text{fb}^{-1}$
is $7000$.
 Branching ratios of the $Z^\prime$ decay
are shown in Table~\ref{table:BRZp}
for Set~A and Set~B\@.
 Similar $\BR(Z^\prime \to XX)$ are predicted 
for the two sets
because $Z^\prime$ is sufficiently heavier than the others.
 The large $\BR(Z^\prime \to \ell\overline{\ell})$
(cf.\ $\BR(Z \to \ell\overline{\ell}) \simeq 0.1$ in the SM)
could be utilized for discovery of the $Z^\prime$ at the LHC\@.
 The $\BL$ nature of the $Z^\prime$ would be tested
if
$\BR(Z^\prime \to b\overline{b})
 /\BR(Z^\prime \to \mu\overline{\mu}) = 1/3$
would be confirmed.

\begin{table}[t]
\begin{center}
\begin{tabular}
{|p{15mm}|p{15mm}|p{15mm}|p{15mm}|p{15mm}|p{15mm}|p{15mm}|p{15mm}|p{15mm}|p{15mm}|}
   \hline
   {}
     & \multicolumn{9}{|c|}{$\BR(Z^\prime \to XX)$}\\
   \hline
    {}
     & $q \overline{q}$
     & $\ell^- \ell^+$
     & $\nu_L^{} \overline{\nu_L^{}}$
     & $\nu_R^{} \overline{\nu_R^{}}$
     & $\Psi_1 \overline{\Psi_1}$
     & $\Psi_2 \overline{\Psi_2}$
     & $s^0_1 (s^0_1)^\ast$
     & $s^0_2 (s^0_2)^\ast$
     & $\eta^+ \eta^-$
    \\ \noalign{\hrule height 1.8pt}
    Set~A
     & $0.20$
     & $0.30$
     & $0.15$
     & $0.10$
     & $0.13$
     & $0.085$
     & $0.012$
     & $0.011$
     & $0.011$
    \\ \hline
    Set~B
     & $0.21$
     & $0.31$
     & $0.16$
     & $0.10$
     & $0.089$
     & $0.089$
     & $0.013$
     & $0.012$
     & $0.012$
    \\ \hline
\end{tabular}
\caption{Branching ratios of $Z^\prime$ decays.}
\label{table:BRZp}
\end{center}
\end{table}

 Since each $U(1)_\DM$-charged particle
decays finally into a dark matter,
$25\,\%$ of the $Z^\prime$
gives a pair of the dark matter~($\Psi_1 \overline{\Psi_1}$)
for Set~A
while $22\,\%$ of the $Z^\prime$
produces $s^0_1 (s^0_1)^\ast$ for Set~B\@.
 Since $f_{\ell i}$ are preferred to be small
in order to satisfy the constraint on $\BR(\meg)$,
$\Psi_2$ for Set~A ($\Psi_1$ and $\Psi_2$ for Set~B)
decays into $\nu_R^{}\, s^0_1$
with $h_{ij}$ and $(y_3)_{ij}$.
 Subsequently
$\nu_R^{}$ dominantly decays into $W^\pm \ell^\mp$ or $Z\nu_L^{}$
(See Table~\ref{table:BRnR}),
and thus it gives a visible signal.
 For Set~A,
$s^0_2$ ($\simeq \eta^0$)
decays invisibly into $\Psi_1 \overline{\nu_L^{}}$ with $f_{ij}$,
and the $s^0_1$ ($\simeq s^0$)
decays also invisibly into
$\Psi_1 \overline{\nu_L^{}}$ with $f_{ij} \cos\theta$.
 For Set~B,
$s^0_2$ decays into $s^0_1 h^0$ with $\mu_3^{}$.
 It is clear that
$\eta^\pm$ provide visible signals
with $\eta^\pm \to \ell^\pm \overline{\Psi_1}$ for Set~A
and $W^\pm s^0_1$ for Set~B\@.
 As a result, about $30\,\%$ ($36\,\%$)
of the $Z^\prime$ for Set~A (B) is invisible
and the constraint on $m_{Z^\prime}$ becomes milder.
 If a light $Z^\prime$ ($\lesssim 1\,\TeV$)
is discovered at the LHC by $Z^\prime \to \ell\overline{\ell}$,
this model could be tested further
at future linear colliders
by measuring the amount of invisible decay of the $Z^\prime$.

\begin{table}[t]
\begin{center}
\begin{tabular}
{|p{15mm}|p{15mm}|p{15mm}|p{15mm}|p{15mm}|}
   \hline
   {}
     & \multicolumn{4}{|c|}{$\BR(\nu_R^{} \to XY)$}\\
   \hline
    {}
     & $W^\pm \ell^\mp$
     & $Z \nu_L^{}$
     & $h^0 \nu_L^{}$
     & $H^0 \nu_L^{}$
    \\ \noalign{\hrule height 1.8pt}
    Set~A
     & $0.53$
     & $0.28$
     & $0.10$
     & $0.09$
    \\ \hline
    Set~B
     & $0.52$
     & $0.28$
     & $0.21$
     & $0$
    \\ \hline
\end{tabular}
\caption{
Branching ratios of $\nu_R^{}$ decays
for Set~A and Set~B
where $(m_R^{})_1 = (m_R^{})_2$ is assumed.
 For Set~B,
the decay $\nu_R^{} \to H^0 \nu_L$ is kinematically forbidden.
}
\label{table:BRnR}
\end{center}
\end{table}

 It is a good feature of this model
that a light $\nu_R^{}$ is acceptable naturally
because of loop-suppressed Dirac mass terms.
 For Set~A,
we see that $Z^\prime \to \nu_R^{} \overline{\nu_R^{}}$
and $Z^\prime \to \Psi_2 \overline{\Psi_2}$
followed by $\Psi_2 \to s^0_2 \nu_R^{}\ (s^0_2 \overline{\nu_R^{}})$
produce about 1200 pairs of $\nu_R^{}$
from $7000$ of $Z^\prime$.
 For Set~B,
the number of $\nu_R^{}$ pairs increases to about 1700
because of an additional contribution from
the decay of $\Psi_1$.
 The $\nu_R^{}$ decays into
$W^\pm \ell^\mp$, $Z \nu_L^{}$,
$h^0 \nu_L^{}$, and $H^0 \nu_L^{}$
through mixing due to the 1-loop induced Dirac Yukawa coupling.
 For Set~A and Set~B,
$\nu_R^{} \to W^\pm \ell^\mp$ is the main decay mode
as shown in Table~\ref{table:BRnR}.
 Then,
the Majorana mass of $\nu_R^{}$ can be reconstructed by
observing the invariant mass of the $jj\ell^\pm$%
~\cite{:1999fq,Ferrari:2002ac}.
 The $h^0$ produced from $\nu_R^{}$ will be energetic
due to the helicity structure.
 Therefore
it would be possible to test the existence of $\nu_R^{}$
by search for energetic $b\overline{b}$.

 If $\Psi_1$ is the dark matter,
the Yukawa coupling constant
$y_{\Psi_1 \Psi_1 h}^{}$~[$= (y_\Psi^{})_1 \sin\alpha$]
for the decay $h^0 \to \Psi_1 \overline{\Psi_1}$
should be small because of small $m_{\Psi_1}^{}$.
 For Set~A,
we have $y_{\Psi_1 \Psi_1 h}^{} \simeq 0.01$.
 Then
main decay mode of $h^0$~$(m_{h^0}^{} < 2 m_t)$ is
$h^0 \to b\overline{b}$ similarly to the SM case.
 Since $\sin\alpha = {\mathcal O}(1)$ is required
to obtain the appropriate relic abundance of $\Psi_1$,
the main decay mode of $H^0$ is also $H^0 \to b\overline{b}$.
 Their decay widths are about a half
of the width of the SM Higgs boson.
 The large mixing prefers
that $m_{h^0}^{}$ and $m_{H^0}^{}$ are of the same order of magnitude.
 Thus
we would find two SM-like Higgs bosons
whose masses are ${\mathcal O}(100)\,\GeV$,
e.g.\ $m_{h^0}^{} = 120\,\GeV$ and $m_{H^0}^{} = 140\,\GeV$ for Set~A\@.

 On the other hand,
if $s^0_1$ is the dark matter,
the interaction of $h^0$ with dark matter $s^0_1$
should satisfy eq.~\eqref{eq:WMAP-B}.
 Then the invisible decay $h^0 \to s^0_1 (s^0_1)^\ast$ dominates
for $2 m_{s^0_1} \leq m_{h^0}^{} < 2 m_W^{}$.
 We have $\BR(h^0 \to s^0_1 (s^0_1)^\ast) = 0.38$ for Set~B\@
where $h^0$ is $100\,\%$ SM-like.
 Therefore,
even if $h^0$ is not discovered at the LHC within a year,
a light $h^0$ might exist
because the constraint on $m_{h^0}^{}$
is relaxed from the one for the SM Higgs boson.
 If such a light $h^0$ is discovered late
with smaller number of signals than the SM expectation,
this model would be confirmed at the LHC~\cite{Eboli:2000ze,invisLHC}
and future linear colliders~\cite{invisILC}
by ``observing'' the $h^0$ invisible decay%
\footnote{
 When $m_{h^0}^{} < 2 m_{s^0_1}$,
the Higgs portal dark matter
such as the $s^0_1$ in this scenario would be
able to be tested at the LHC~\cite{Kanemura:2010sh}
and future linear colliders~\cite{Kanemura:2011nm}.
}.

\subsection{Some remarks}
\label{subsec:remarks}

 In our paper,
we have assumed the mass of the $Z'$ boson to be $2\,\TeV$.
 It is expected that
the lower bound on the mass will be more and more stringent
due to the new results from the LHC
and that the above value of the mass would be excluded in near future.
 In such a case,
the mass should be taken to be higher values than $2\,\TeV$.
Then,
the branching ratio of $Z' \to \Psi_2 \overline{\Psi}_2$
becomes closer values to that of $Z' \to \Psi_1 \overline{\Psi}_1$
because the effect of their mass difference
becomes less significant.
 Heavy $Z'$ is achieved
by assuming larger values of $g_\BL^{}$
or assuming larger values of $v_\sigma^{}$.
 For the former case,
most of the phenomenological analyses are unchanged.
 For the latter case,
our phenomenological analysis would be changed slightly.
 Even in this case,
the experimental constraints from neutrino experiments
and the $\mu\to e\gamma$ results
can be satisfied
by using smaller values of $(y_R^{})_i$ and $(y_\Psi^{})_i$
which keep $(m_R^{})_i$ and $m_{\Psi_i}^{}$ unchanged
from values in our Sets~A and B\@.
 In the scenario of Set~A,
a smaller value of $(y_\Psi^{})_1$
results in a larger value of the DM abundance
which is proportional to $(y_\Psi^{})_1^{-2}$.
 Even if the red curve in Fig.~\ref{fig:WMAP}(a)
goes up by about a factor of 10,
the WMAP result can still be explained
by $m_{\Psi_1}^{}\simeq 60\,\GeV$.
 Therefore,
we can accept three times larger $v_\sigma$
(namely, three times smaller $(y_\Psi^{})_1$)
than the value~($5\,\TeV$) we used.

 A shortage is about the gauge anomaly for $U(1)_\BL$.
 It is well known that
$U(1)_\BL$ is free from anomaly
if three singlet fermions of $\BL=-1$
(usual right-handed neutrinos)
are added to the SM\@.
 However,
we have introduced not three but only two $(\nu_R^{})_i$,
and their $\BL$ is not $-1$ but $+1$.
 There are also extra $U(1)_\BL$-charged fermions
($(\Psi_L)_i$ and $(\Psi_R)_i$).
 Therefore
the $U(1)_\BL$ gauge symmetry has anomalies
for the triangle diagrams
of $[U(1)_\BL]^3$ and $[U(1)_\BL]\times [\text{gravity}]^2$.
 They would be resolved
by some heavy singlet fermions
with appropriate $\BL$~(See e.g., ref.~\cite{anomaly-cancel}).

 Another is
the way to reproduce the baryon asymmetry of the universe.
 Since we have used only particles below the TeV-scale,
leptogenesis~\cite{Fukugita:1986hr} does not work
in a natural way.
 Heavy fermions to eliminate the $U(1)_\BL$ gauge anomaly might help.
 The electroweak baryogenesis~\cite{Kuzmin:1985mm}
would be accommodated
by the introduction
of an additional Higgs doublet to the model
so that new source of CP-violation
appears in the Higgs potential.

\section{Conclusions}
\label{sec:concl}

 We have presented the TeV-scale seesaw model
in which $U(1)_\BL$ gauge symmetry
can be the common origin
of neutrino masses,
the dark matter mass (if $\Psi_1$ is the one),
and stability of the dark matter.
 Light neutrino masses are generated by the two-loop diagrams
which are also contributed by the dark matter,
a Dirac fermion $\Psi_1$ or a complex scalar $s^0_1$.
 The symmetry to stabilize the dark matter
appears in the $U(1)_\BL$-symmetric Lagrangian
without introducing additional global symmetry
(ex.\ a $Z_2$ symmetry).
 It has been shown that
this model can be compatible with constraints
from the neutrino oscillation data,
the search for $\meg$,
the relic abundance of the dark matter,
and the direct search results for dark matter.
 It should be emphasized that
these constraints are satisfied
with sizable coupling constants ($\gtrsim 10^{-2}$)
and new particles (including $\nu_R^{}$)
whose masses are at or below the TeV-scale.

 We have discussed collider phenomenology in this model
by using two sets of parameters
which satisfy constraints from the current experimental data.
 The $U(1)_\BL$ gauge boson $Z^\prime$
can be discovered at the LHC
by observing $Z^\prime \to \ell\overline{\ell}$
if it is not too heavy.
 In our model,
since the dark matter has $U(1)_\BL$ charge,
$Z^\prime$ partially decays into a pair of the dark matter.
 As a result,
more than $30\,\%$ of produced $Z^\prime$
is invisible for both the sets.
 Then
a lighter $Z^\prime$ is allowed than
the usual experimental bound.
 Detailed studies for such a $Z^\prime$
would be performed
at future linear colliders.

 Since masses of $(\nu_R^{})_i$ and $\Psi_i$
are obtained by the $U(1)_\BL$ breaking,
it would be natural that
$\Psi_1$ is light and becomes the dark matter
when we assume $\nu_R^{}$ masses are of ${\mathcal O}(100)\,\GeV$.
 In this case,
a large mixing between $h^0$ and $H^0$ is required
for the appropriate relic abundance
because the Yukawa coupling constant is small.
 Decay branching ratios of $h^0$ and $H^0$
are almost the same as the one for the SM Higgs boson.
 Therefore
two SM-like Higgs bosons with similar masses
(ex.\ $m_{h^0}^{} = 120\,\GeV$ and $m_{H^0}^{} = 140\,\GeV$ for Set~A)
would be discovered at the LHC\@.

 On the other hand,
if $s^0_1$ is the dark matter,
the decay of the lighter Higgs boson $h^0$
can be dominated
by invisible $h^0 \to s^0_1 (s^0_1)^\ast$.
 For Set~B,
we obtain $\BR(h^0 \to s^0_1 (s^0_1)^\ast) = 38\,\%$.
 Then
this model would be tested
at future linear colliders
by measuring the amount of the invisible decay.

\begin{acknowledgments}
 This work was supported in part
by Grant-in-Aid for Scientific Research,
Japan Society for Promotion of Science (JSPS),
Nos.\ 22244031 and 23104006 (S.K.)
and No.\ 23740210 (H.S.).
 The work of H.S.\ was supported
by the Sasakawa Scientific Research Grant
from the Japan Science Society.
\end{acknowledgments}

\end{document}